\documentclass[9pt,twocolumn]{article}
\usepackage{lineno,amsmath,graphicx,hyperref}
\usepackage{mathtools}
\usepackage{amssymb} 
\usepackage{lmodern}
\usepackage{geometry}
\usepackage{booktabs}
\usepackage{adjustbox}
\newcommand{\Tr}{\operatorname{Tr}}

\usepackage[numbers]{natbib}  

\makeatletter
\newenvironment{revtexabstract}
{\par\vspace{10pt}\centering%
	\begin{minipage}{0.9\linewidth}%
		\centerline{\large\bfseries Abstract}\vspace{0.5em}\small\noindent\ignorespaces}
	{\end{minipage}\par\vspace{10pt}}
\makeatother

\begin{document}

\title{The formation of entangled Schr\"odinger cat-like states in the process of spontaneous parametric down-conversion}

\author{
	R. Singh\thanks{ranjit.singh@mail.ru} \\
	\small Independent Researcher, Domodedovo, 142000, Moscow region, Russia \\
	A.E. Teretenkov\thanks{taemsu@mail.ru} \\
	\small Department of Mathematical Methods for Quantum Technologies, \\
	\small Steklov Mathematical Institute of Russian Academy of Sciences, \\
	\small 8 Gubkina St., Moscow, 119991, Russia
}

\date{\today}

	\maketitle

\begin{revtexabstract}
		We investigate entangled Schr\"odinger cat-like states (SCLSs) in degenerate and non-degenerate spontaneous parametric down-conversion (SPDC) with a fully quantized, depleted pump. Our fully quantum treatment, visualized via Wigner functions, reveals non-Gaussian features and interference patterns absent in semiclassical models. For degenerate SPDC, we demonstrate significant squeezing (up to $4.04\,\mathrm{dB}$) and robust super-Poissonian statistics in both non-dissipative and dissipative regimes. Extending to non-degenerate SPDC, we confirm that pump quantization also generates non-Gaussian states in all modes and yields a higher-dimensional entanglement structure, evidenced by a larger Schmidt number ($K^{(\mathrm{ND})} \approx 10.38$) compared to the degenerate case ($K \approx 1.93$). Our approach captures critical dynamics like energy exchange and phase-dependent evolution. These entangled SCLSs, non-Gaussian states realizable in $\chi^{(2)}$ media at moderate intensities and offering advantages over $\chi^{(3)}$-based schemes, are promising resources for quantum sensing and information processing.
	\end{revtexabstract}
	
	
	\section{Introduction}
	\label{introduction}
	
	We investigate the formation of entangled Schr\"odinger cat-like states (SCLSs), non-Gaussian states in the fundamental and second-harmonic modes of spontaneous parametric down-conversion (SPDC) under a fully quantized pump model. Unlike conventional optical Schr\"odinger cat states (even/odd coherent states \cite{Dodonov1974}) that exhibit only non-classical interference, our SCLSs demonstrate additional quantum features including significant squeezing (up to $4.04\,\mathrm{dB}$) and super-Poissonian statistics (Fano factor $\gg 1$), both robust under moderate losses. These properties, combined with their inherent entanglement between modes, distinguish them fundamentally from single-mode cat states. Going beyond undepleted-pump approximations, we analyze their dynamics in non-dissipative and dissipative regimes. Wigner function visualizations reveal the non-Gaussian character and interference patterns of these states, confirming their SCLS nature. Unlike semiclassical models that produce statistical mixtures \cite{w1988,Walls2008}, our approach captures energy exchange, phase-dependent evolution, and interference, highlighting critical properties for continuous-variable quantum technologies \cite{Braunstein2005}.
	
	These entangled SCLSs, non-Gaussian states are promising for quantum sensing and information processing, offering a path to strong non-Gaussianity in $\chi^{(2)}$ media at moderate pump intensities. Furthermore, we extend our investigation to the non-degenerate (NDSPDC) configuration, demonstrating that pump quantization similarly generates non-Gaussian resources and yields richer, tripartite entanglement structures compared to the degenerate case.
	
	While the standard treatment of SPDC approximates the Hamiltonian as quadratic by treating the pump classically, our work employs the full cubic Hamiltonian \(\hat{H}_{\mathrm{int}} \propto \hat{a}_1^2 \hat{a}_2^{\dagger} + \mathrm{h.c.}\) or $\hat{H}_{\mathrm{int}}^{(\mathrm{ND})} \propto \hat{a}_s \hat{a}_i \hat{a}_p^{\dagger} + \mathrm{h.c.}$ with a quantized pump. The key novelty lies in exploring the high-depletion operating condition---where specific initial states lead to approximately 63.2\% pump conversion (for degenerate SPDC process) --- whose quantum features had remained largely uncharacterized due to the dominance of the undepleted-pump approximation. This operating condition could be realized with modern highly-efficient nonlinear crystals (e.g., periodically poled structures \cite[see][and references therein]{Belinsky2018}) and using contemporary numerical tools like QuTiP \cite{johansson2012qutip,JOHANSSON20131234}.
	
	SCLSs, characterized by negative values of the Wigner function, play an important role in, e.g., observing phase displacements due to their phase sensitivity \cite{Shukla2023,Toscano2006,Salykina2023,Singh2024}. Schr\"odinger cat states can be used to encode cat qubits \cite{Cochrane1999} and building Ising machines \cite{Yamamoto2020}.
	
	In the field of quantum nonlinear optics, the generation of non-Gaussian states requires at least the cubic form of the interaction Hamiltonian \cite{Braunstein2005}. The formation of non-Gaussian states \cite{Sychev2017} (e.g., SCLSs) has been considered in \cite{Nikitin1991,Singh2021} during quantum state evolution in second harmonic generation with $\chi^{(2)}$ media, with \cite{Singh2021} demonstrating interference patterns using Wigner functions.
	
	The classical and non-depleted or semi-classical method approximates the interaction Hamiltonian for the parametric down-conversion (PDC) process, which is based on $\chi^{(2)}$ (and is cubic), by a quadratic one (i.e., pump mode is considered classic and undepleted). Such an approximation neglects the non-Gaussianity features present in the initial cubic form of the interaction Hamiltonian. Another approximation method, e.g., the expansion of unitary operators \cite{Agarwal1974,Beskrovnyi1996, Belinsky2018}, can be applied to consider a full treatment of quantum effects present in all the modes of the PDC process, but starts to deviate at later stages \cite{Belinsky2018}. However, the diagonalization method \cite{Walls1970,Walls1972,Gantsog1991,Tanas1992,Belinsky2018} can be used to study evolution at longer interaction lengths without losing the quantum effects present in the interacting modes.
	
	Previous studies implemented fully quantum mechanical treatments of parametric down-conversion (PDC) with pump depletion, explicitly accounting for the correlated exchange of quanta between the pump and down-converted modes \cite{Walls1970,Walls1972,Gantsog1991,Tanas1991,Tanas1992,Drobny1992}. These works investigated statistical properties such as mean photon numbers, quadrature fluctuations in the fundamental mode, and phase properties of both the pump and signal fields \cite{Gantsog1993}. 
	
	Subsequent theoretical analyses clarified the limitations of the parametric approximation by demonstrating that its validity is governed not by the absence of pump depletion, but by the preservation of pump coherence after the interaction \cite{Dariano1999}. Extensions of this insight to experimentally relevant regimes examined the role of quantum fluctuations, non-Gaussian state generation, and critical behavior near threshold in optical parametric oscillators \cite{DAuria2010}. More recent studies have further explored pump depletion, entanglement, and squeezing dynamics beyond the parametric approximation in macroscopic down-conversion processes, identifying the characteristic timescales over which genuinely quantum features emerge simultaneously in the pump and down-converted modes \cite{Chinni2023}.
	
	Semiclassical treatments of PDC~\cite{w1988,Walls2008} assume a classical, undepleted pump field, leading to a Fokker--Planck description whose steady-state solution corresponds to a statistical mixture of two coherent states. While such approaches capture mean-field amplification and phase bistability, this description inherently misses essential quantum features: it exhibits no interference structure, cannot represent genuine even or odd Schr\"odinger-cat states, neglects energy exchange and back-action between modes, shows no dependence on interaction length, and remains confined to a dissipative steady-state regime.
	
	Our full quantum treatment overcomes these limitations, demonstrating the formation of SCLSs in both modes across both non-dissipative and dissipative regimes. Crucially, our results preserve the quantum interference structure and provide direct evidence of energy exchange between the fundamental and second-harmonic modes (see Fig.~\ref{fig:fig1} and Fig.~\ref{fig:fig5}).
	
	In this paper, we have studied the formation of entangled SCLSs for two cases (non-dissipative and dissipative) when both modes (fundamental and second harmonic) are considered as quantum and depleted during the effective realization of the degenerate SPDC process based on $\chi^{(2)}$ \cite{Walls2008,Agarwal2013}. We extend this methodology to the non-degenerate SPDC (NDSPDC) configuration, analyzing the formation of non-Gaussian states and entanglement in its tripartite system under the same fully quantum framework. QuTiP \cite{johansson2012qutip,JOHANSSON20131234} is used to numerically solve the Lindbladian superoperator responsible for the SPDC process. The formation of SCLSs is studied and illustrated qualitatively using the Wigner function. For the degenerate SPDC process, quantitative analysis of the SCLSs is performed by calculating squeezing levels (variances of quadrature components and Fano factor), values of photon number distributions, and the entanglement between the two modes. The SCLSs are observed in the fundamental and second harmonic modes for non-dissipative and dissipative cases when the fundamental mode is in the vacuum state and the second harmonic mode is in the coherent state. For the NDSPDC process, we similarly compute Wigner functions and quantify entanglement via Schmidt number analysis. In this configuration, we observe the formation of non-Gaussian states: the signal mode exhibits a non-Gaussian structure distinct from a thermal state, while the pump mode develops SCLS features.
	
	\section{Degenerate SPDC under pump depletion and losses}
	
	Let three stationary degenerate monochromatic optical plane wave modes $\hat{a}_s,\hat{a}_i,\hat{a}_p$ of frequencies $\omega_s$, $\omega_i$, $\omega_p$ propagate collinearly in a nonlinear optical crystal with non-zero susceptibility $\chi^{(2)}$.  Subscripts $i,s,p$ belong to the idler, signal, and pump modes. In case of degenerate frequencies: $\omega_s=\omega_i=\omega$, $\omega_p=2\omega$, $\hat{a}_i =\hat{a}_s = \hat{a}_1, \hat{a}_p=\hat{a}_2$. For effective realization of the SPDC process, it is assumed that all three modes can be phase-matched \cite{Dmitriev1999} or quasi-phase-matched \cite{Belinsky2018}. The interaction Hamiltonian for the degenerate SPDC process $(2\omega = \omega + \omega)$ \cite{Dariano1999,Walls2008,Agarwal2013}
	
	\begin{equation}
		\hat{H}_{int} = \hbar g (\hat{a}^{2}_1 \hat{a}_2^{\dagger} + \mathrm{h.c.}),
		\label{eq:hint}
	\end{equation}
	where $\hbar$ is the Planck constant. For simplicity $\hbar=1$. $g$ is the nonlinear coupling constant of the interacting modes. 
	
	The Lindblad master equation for the density matrix $\hat{\rho}$ describing an open quantum system is used to study the quantum dynamics of the interaction Hamiltonian (\ref{eq:hint}) and can be written as \cite{Agarwal2013, Walls2008, BP2002, Strinati2024} 
	\begin{equation}
		\frac{d \hat{\rho}}{d \tau} = -ig^{-1}[\hat{H}_{int}, \hat{\rho}] + \sum_{j=1}^{2}\left(\hat{C}_j \hat{\rho} \hat{C}_j^{\dagger} - \frac12\{ \hat{C}_j^{\dagger}\hat{C}_j, \hat{\rho}\}\right),
		\label{eq:lind}
	\end{equation}
	where $\tau = g t$ is the normalized interaction length.  $\hat{C}_j = \sqrt{\gamma_j/g} \hat{a}_j$  are the Lindblad operators describing the dissipative part of the dynamics, where $\gamma _{j}\geq 0$  are the cavity damping rates of the modes and  $g> 0$.

	Equation (\ref{eq:lind}) is numerically solved by using QuTiP for two cases (non-dissipative $\gamma_j =0$ and dissipative $\gamma_j = 0.15$) for the initial state density matrix $\hat{\rho}(0) = | \psi_0 \rangle \langle \psi_0|$. At the input of the nonlinear crystal, the fundamental mode is in the vacuum state $|0 \rangle_1$, the second harmonic one is in the coherent state $|\alpha_{20}\rangle_2 = e^{-|\alpha_{20}|^2/2} \sum_{n_2=0}^{\infty} \frac{\alpha_{20}^{n_2}}{\sqrt{n_2!}}| n_2 \rangle$, having mean number of photons $|\alpha_{20}|^2=20$ and phase $\varphi_{20}=\pi/2$, i.e.,  $| \psi_0 \rangle = |0\rangle_1 \otimes |\alpha_{20} \rangle_2$.
	
	\section{Formation of cat-like states in both modes}
	One can study and analyze the quantum statistical properties of the modes by using the Wigner quasiprobability distribution \cite{Walls2008,Agarwal2013}. Phase space portraits of the Wigner function help to visualize delicate patterns, such as (a) interference (wave nature) present in the superposition of macroscopically distinct states, e.g., SCLSs, (b) non-Gaussianity of the state, i.e., the negative value of the Wigner function. The Wigner functions of the modes $\hat{a}_1$ and $\hat{a}_2$ are calculated using  
	
	\begin{equation}
		W_{j}(\alpha_j,\tau) = \frac{1}{\pi} \int_{-\infty}^{\infty} \langle \Re{(\alpha_j)}-y| \hat{\rho}_j(\tau) |\Re{(\alpha_j)}+y \rangle e^{2 i \Im{(\alpha_j)} y} dy,
		\label{eq:wigner1}
	\end{equation}
	where $j=1,2$, $\hat{\rho}_1(\tau) = \Tr_{2} [\hat{\rho}(\tau)]$, $\hat{\rho}_2(\tau) = \Tr_{1} [\hat{\rho}(\tau)]$ are reduced density matrices of states of the modes $\hat{a}_1$ and $\hat{a}_2$.

	\begin{figure}[h!]
		\centering
		\includegraphics[width=1\linewidth]{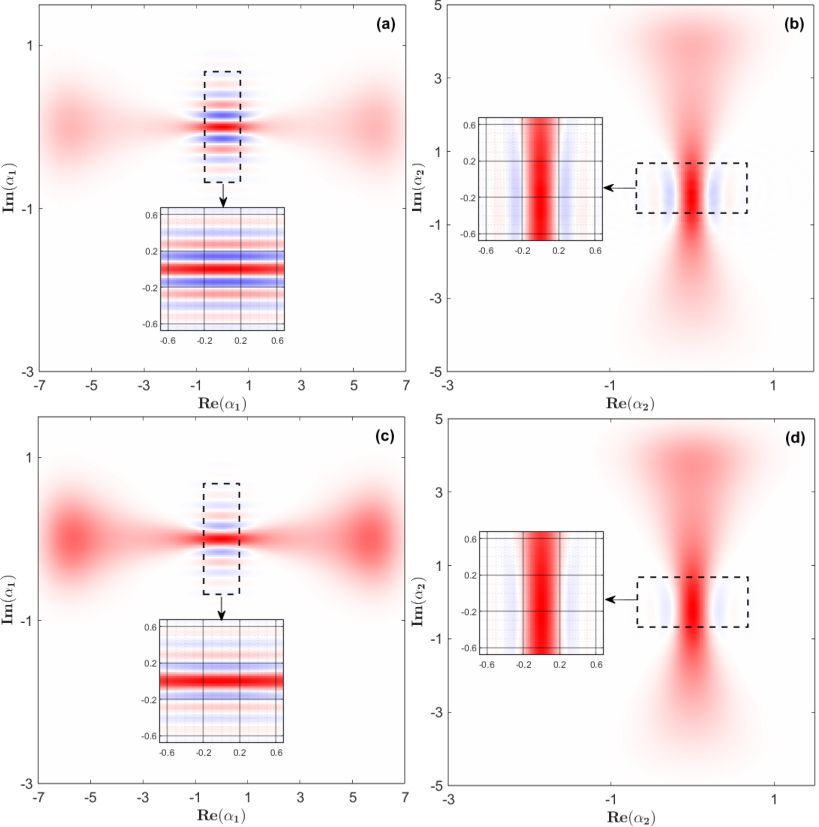}
		\caption{Wigner functions $W_1(\alpha_1,\tau)$ and $W_2(\alpha_2,\tau)$ for degenerate SPDC process at $\tau = 0.38$. Top row (a,b): Non-dissipative ($\gamma_{1,2} = 0$), $N_1 \approx 25.28$, $N_2 \approx 7.35$. Bottom row (c,d): Dissipative ($\gamma_{1,2} = 0.15$), $N_1 \approx 23.88$, $N_2 \approx 6.94$. Initial: $\hat{a}_1$ vacuum, $\hat{a}_2$ coherent state ($|\alpha_{20}|^2 = 20$, $\varphi_{20} = \pi/2$).}
		\label{fig:fig1}
	\end{figure}

	In Fig.~\ref{fig:fig1} plots of (\ref{eq:wigner1}) are shown for two cases (non-dissipative and dissipative) of the formation of SCLSs in the fundamental and second harmonic modes.
	
	\section{Application-relevant statistical properties}
	Using the Wigner function, we calculated the contribution of all statistical moments present in the SCLSs and qualitatively visualized portraits of the Wigner functions in phase space (see Fig.~\ref{fig:fig1}). In order to quantitatively estimate quantum statistical properties such as squeezing level (the variances of quadrature components, the Fano factor) photon number distribution calculations and the entanglement measure are performed. 
	
	\subsection{Mean number extrema as indicators of SCLSs formation. Squeezing in both modes}
	
	The mean number of photons in the both modes $\hat{a}_j$ is calculated using
	\begin{eqnarray}
		N_j (\tau)= \Tr [\hat{a}^{\dagger}_j \hat{a}_j \hat{\rho}_j(\tau)]. \label{eq:nj}
	\end{eqnarray}

\begin{figure}[h!]
		\centering
		\includegraphics[width=0.9\linewidth]{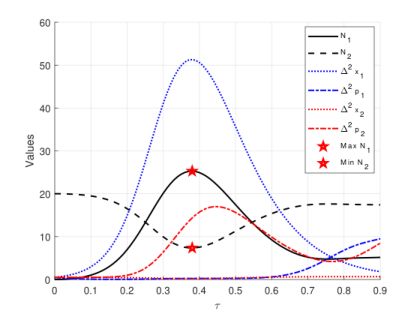}
		\caption{Evolution of (a)~$N_j$ and (b)~quadrature variances for degenerate SPDC process. $N_1$ ($\hat{a}_1$, solid black, $-$) peaks while $N_2$ ($\hat{a}_2$, dashed black, $--$) minimizes at $\tau=0.38$. Variances: $\Delta^2x_1$ (blue $\cdot{\cdot}{\cdot}$), $\Delta^2p_1$ (blue $\cdot$-), $\Delta^2x_2$ (red $\cdot{\cdot}{\cdot}$), $\Delta^2p_2$ (red $\cdot$-). $\bigstar$ marks extrema.}
		\label{fig:fig5}
	\end{figure}

	The variances of the quadrature components are calculated using
	\begin{eqnarray}
		\Delta^2 x_j = \Tr [\hat{x}_j^2 \hat{\rho}_j(\tau)] - \Tr [ \hat{x}_j \hat{\rho}_j(\tau)]^2, \label{eq:quad1}\\
		\Delta^2 p_j = \Tr [\hat{p}_j^2 \hat{\rho}_j(\tau)] - \Tr [ \hat{p}_j \hat{\rho}_j(\tau)]^2, \label{eq:quad2}
	\end{eqnarray}
	where $\hat {x}_j = 2^{-(1/2)}(\hat{a}_j + \hat{a}^{\dagger}_j) $ and $\hat {p}_j = -i2^{-(1/2)}(\hat{a}_j - \hat{a}^{\dagger}_j) $ are the quadrature components of the modes $\hat{a}_j$. 
	Fig.~\ref{fig:fig5} shows the evolution of both (i) the mean photon numbers (Eq.~\ref{eq:nj}) and (ii) the quadrature component variances (Eqs.~\ref{eq:quad1} and~\ref{eq:quad2}) for the non-dissipative case. SCLSs form at the first extremal values of the photon numbers: the maximum for mode $\hat{a}_1$ and the minimum for mode $\hat{a}_2$, both occurring at $\tau=0.38$. Notably, the quadrature variances $\Delta^2 p_1(0.38) \approx 0.1976$ (mode $\hat{a}_1$) and $\Delta^2 x_2(0.38) \approx 0.1970$ (mode $\hat{a}_2$) demonstrate squeezing, corresponding to $4.03$ dB and $4.04$ dB respectively.
	
	\begin{figure}[h!]
		\centering
		\includegraphics[width=1\linewidth]{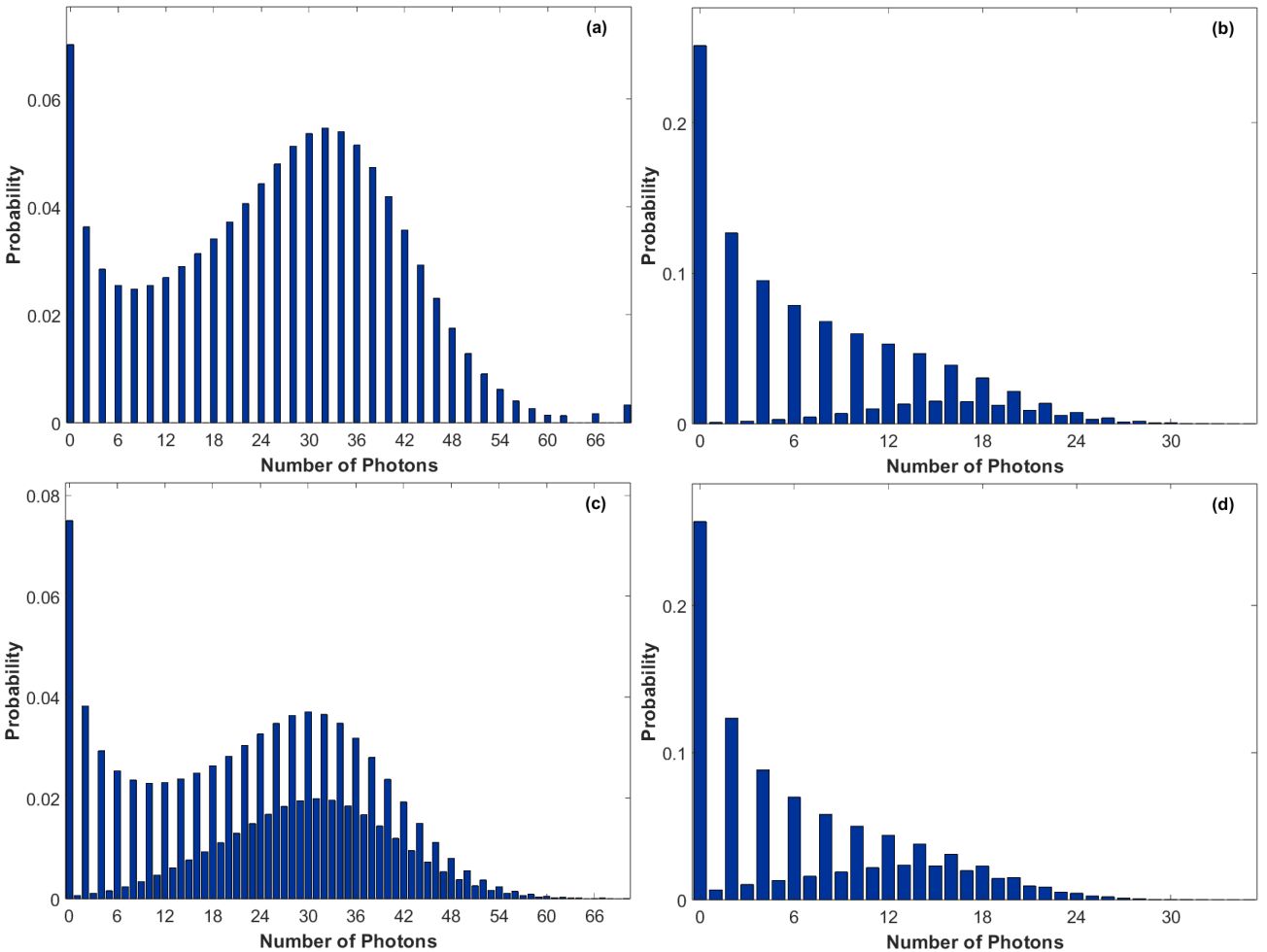}
		\caption{Photon number distributions $P_j(n_j)$ at $\tau = 0.38$ for degenerate SPDC process: (a,b) non-dissipative ($\gamma_{1,2}=0$) and (c,d) dissipative ($\gamma_{1,2}=0.15$) regimes for modes $\hat{a}_1$ (a,c) and $\hat{a}_2$ (b,d).}
		\label{fig:fig3}
	\end{figure}
	
	\subsection{Dissipation-induced parity mixing in SCLSs}
	It helps to identify the type of pattern an SCLS belongs to, e.g. even, odd, mixture of even and odd coherent states. The photon number distribution is calculated using	
	\begin{eqnarray}
		P_j(n_j) = \Tr  \left[ \hat{\rho}_j {|n_j\rangle}{\langle n_j|} \right]. \label{eq:pnd}	
	\end{eqnarray}
	Fig.~\ref{fig:fig3} shows pattern types of photon number distributions (\ref{eq:pnd}) of SCLSs $\hat{\rho}_j$. In case of non-dissipative regime, photon number distributions of SCLSs $\hat{\rho}_j$ for modes $\hat{a}_1$ and $\hat{a}_2$ are associated with even and mixture of even and odd coherent states. In the case of the dissipative regime, Figs.~\ref{fig:fig3}(c)-\ref{fig:fig3}(d) show that the SCLSs $\hat{\rho}_j(\tau)$ are associated with the photon number distribution of the mixture of even and odd coherent states. The dissipative regime introduces an odd number of photons into the SCLSs $\hat{\rho}_j(\tau)$ (see Figs.~\ref{fig:fig3}(c)-\ref{fig:fig3}(d)).

	\subsection{Robustness  of super-Poissonian statistics to dissipation}
	The value of the Fano factor can identify the type (sub-Poissonian, Poissonian, super-Poissonian) of the distributions of the studied SCLS $\hat{\rho}_j$. Thus, the Fano factor \cite{Vogel} is calculated using
	\begin{eqnarray}
		FF_j (\tau) = \frac{\Tr [ (\hat{a}^{\dagger}_j \hat{a}_j)^2 \hat{\rho}_j(\tau)] - N_j^2(\tau) } {N_j(\tau)}. \label{eq:ff}	
	\end{eqnarray}
	The Fano factor $FF_j$ characterizes photon statistics: $FF_j < 1$ indicates sub-Poissonian, $FF_j = 1$ Poissonian, and $FF_j > 1$ super-Poissonian behavior. At $\tau = 0.38$, the calculated values from Eq.~(\ref{eq:ff}) are: $FF_1 \approx 8.54$ (non-diss.), $8.35$ (diss.); $FF_2 \approx 6.66$ (non-diss.), $6.38$ (diss.)---all indicating super-Poissonian statistics.
	
	\subsection{Entanglement of SCLSs}
	To quantify the entanglement between the two modes $(\hat{a}_1,\hat{a}_2)$ in the SCLSs, we compute the Schmidt number \cite{Law2004} defined as:
	\begin{eqnarray}
		K (\tau) = \frac{1}{\Tr [ (\hat{\rho}_1(\tau))^2]} = \frac{1}{\Tr [ (\hat{\rho}_2(\tau))^2]}, \label{eq:ent}	
	\end{eqnarray}
	where $\hat{\rho}_1$ and $\hat{\rho}_2$ are the reduced density matrices of modes $\hat{a}_1$ and $\hat{a}_2$, respectively. For pure bipartite states, the parameter $K$ serves as an entanglement measure, with $K > 1$ indicating non-separable (entangled) states. At $\tau = 0.38$, the value calculated from Eq.~(\ref{eq:ent}) is $K \approx 1.93$ for the non-dissipative case, confirming the presence of entanglement.
	
	\section{Non-degenerate SPDC: An Extended Analysis}
	\label{sec:ndspdc}
	
	While the degenerate SPDC process is foundational, the non-degenerate (NDSPDC) configuration---where a pump photon splits into two distinguishable photons---is of paramount importance for quantum information processing, enabling heralded single-photon generation and discrete-variable entanglement. In this section, we adapt our fully quantum numerical methodology to the NDSPDC process to investigate the formation and properties of non-Gaussian states and entanglement in contrast to the degenerate case.
	
	\subsection{Hamiltonian and System Description}
	The interaction Hamiltonian for the NDSPDC process, in which a pump photon splits into two distinguishable signal and idler photons ($\omega_p = \omega_s + \omega_i$) propagating collinearly in a $\chi^{(2)}$ nonlinear crystal, is given by~ \cite{Gantsog1993,Dariano1999,Walls2008,DAuria2010,Chinni2023}
	\begin{equation}
		\hat{H}_{\mathrm{int}}^{(\mathrm{ND})} = \hbar g^{(\mathrm{ND})} (\hat{a}_s \hat{a}_i \hat{a}_p^{\dagger} + \mathrm{h.c.}),
		\label{eq:nonDegHint}
	\end{equation}
	where $\hat{a}_p$, $\hat{a}_s$, and $\hat{a}_i$ are the annihilation operators for the pump, signal, and idler modes at frequencies $\omega_p$, $\omega_s$, and $\omega_i$, respectively. We set $\hbar=1$ for simplicity. The nonlinear coupling constant $g^{(\mathrm{ND})}$ differs from its degenerate counterpart $g^{(\mathrm{D})}$ due to distinct phase-matching conditions inherent to the two processes. The open-system dynamics are governed by a Lindblad master equation of the same form as Eq.~(\ref{eq:lind}), with $\hat{H}_{\mathrm{int}}$ replaced by $\hat{H}_{\mathrm{int}}^{(\mathrm{ND})}$ and with dissipation channels for all three modes.
	
	For this brief comparative study, we restrict the NDSPDC analysis to the non-dissipative regime ($\gamma_j=0$) to examine the ideal, lossless quantum dynamics. We choose an initial state analogous to the degenerate case: the signal and idler modes in the vacuum state and the pump mode in a coherent state,
	\begin{equation}
		|\psi_0^{(\mathrm{ND})}\rangle = |0\rangle_s \otimes |0\rangle_i \otimes |\alpha_{p0}\rangle_p,
	\end{equation}
	with $|\alpha_{p0}|^2 = 20$ and phase $\varphi_{p0} = \pi/2$.
	
	\subsection{Mean number of photons and formation of Non-Gaussian States}
	
\begin{figure}[h!]
		\centering\includegraphics[width=1\linewidth]{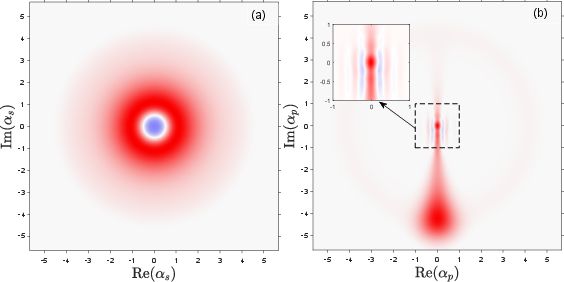}
		\caption{Wigner functions (a) $W_s^{ND}(\alpha_s,\tau)$ and (b) $W_p^{ND}(\alpha_p,\tau)$ for non-degenerate SPDC process at $\tau = 1$, shown for the non-dissipative case ($\gamma_{1,2,3} = 0$). Mean photon numbers: $N_s \approx 5.44$, $N_p \approx 14.55$. Initial conditions: $\hat{a}_s$ and $\hat{a}_i$ in vacuum, $\hat{a}_p$ in a coherent state ($|\alpha_{p0}|^2 = 20$, $\varphi_{p0} = \pi/2$).}
		\label{fig:fig4ab}
	\end{figure}

	We compute the mean photon numbers for both the signal ($\hat{a}_s$) and pump ($\hat{a}_p$) modes using Eq.~(\ref{eq:nj}), and their corresponding Wigner functions $W_s^{(\mathrm{ND})}(\alpha_s,\tau)$ and $W_p^{(\mathrm{ND})}(\alpha_p,\tau)$ using Eq.~(\ref{eq:wigner1}) from the reduced density matrices $\hat{\rho}_s^{(\mathrm{ND})}(\tau)$ and $\hat{\rho}_p^{(\mathrm{ND})}(\tau)$. The resulting phase-space portraits, shown in Figs.~\ref{fig:fig4ab}(a) and~\ref{fig:fig4ab}(b), reveal the formation of non-Gaussian quantum states.
	
	The pump mode's Wigner function [Fig.~\ref{fig:fig4ab}(b)] exhibits clear interference fringes and negative-value regions, a definitive signature of Schr\"odinger cat-like state (SCLS) formation. This parallels the behavior observed for the pump mode in the degenerate SPDC process, confirming that a fully quantized pump leads to non-Gaussian state generation in both $\chi^{(2)}$ processes.
	
	In contrast to the pump, the signal mode's Wigner function [Fig.~\ref{fig:fig4ab}(a)] displays a markedly different structure. The semiclassical, undepleted-pump approximation for SPDC predicts the signal mode to evolve into a thermal (Gaussian) state. However, under the full quantum treatment with pump depletion, we observe the formation of a non-Gaussian state. This deviation thus demonstrates that a fully quantized pump model is required to produce non-Gaussian states in the down-converted fields. Due to the symmetry of the Hamiltonian in Eq.~(\ref{eq:nonDegHint}), the idler mode's Wigner function $W_i^{(\mathrm{ND})}(\alpha_i,\tau)$ is qualitatively identical to that of the signal mode; therefore, its explicit analysis is omitted.
	
	\subsection{Entanglement Structure in NDSPDC}
	
	A fundamental distinction from the two-mode degenerate SPDC process is the tripartite nature of the state generated by NDSPDC. To quantify the bipartite entanglement within this system, we compute the Schmidt number for a specific partition. The entanglement between the signal mode and the combined idler-pump subsystem is
	\begin{equation}
		K^{(\mathrm{ND})} (\tau) = \frac{1}{\Tr [ (\hat{\rho}_s^{(\mathrm{ND})}(\tau))^2 ]} = \frac{1}{\Tr [ (\hat{\rho}_{ip}^{(\mathrm{ND})}(\tau))^2 ]},
		\label{eq:entND}
	\end{equation}
	where $\hat{\rho}_{ip}^{(\mathrm{ND})}$ is the reduced density matrix for the idler and pump modes. For the specified initial state and parameters, we find $K^{(\mathrm{ND})} \approx 10.38$ at $\tau = 1.0$ in the non-dissipative case, indicating strong entanglement. This value is notably larger than the Schmidt number $K \approx 1.93$ found for the bipartite entanglement in the degenerate SPDC process at its characteristic time ($\tau=0.38$), reflecting the higher-dimensional entanglement structure accessible in the non-degenerate system.
	
	\section{Numerical implementation and convergence}
	
	To ensure the reliability and consistency of the numerical simulations, particular care was taken in the truncation of the Hilbert space and in monitoring several internal consistency checks. Since the underlying Hilbert space is infinite dimensional, all numerical calculations were performed in a truncated Fock basis with a sufficiently large cutoff $N_{\mathrm{max}}$. The value of $N_{\mathrm{max}}$ was increased until all relevant physical observables converged and became insensitive to further increases of the cutoff.
	
	As consistency checks, we systematically monitored the following quantities throughout the evolution: 
	(i) the trace of the density matrix, which was verified to remain equal to unity within numerical precision; 
	(ii) the purity in the case of unitary dynamics, quantified by $\mathrm{Tr}(\rho^2)$, which equals one for the full (pure) density matrix and satisfies $\mathrm{Tr}(\rho^2)\leq 1$ for reduced density matrices obtained via partial tracing; 
	(iii) the positivity of the density matrix, verified by confirming the non-negativity of its eigenvalues; 
	(iv) the mean photon numbers and their variances, which were checked for convergence with respect to the Hilbert-space cutoff; and 
	(v) the stability of phase-space representations, in particular the Wigner functions, whose qualitative features and interference patterns were found to be unchanged upon increasing $N_{\mathrm{max}}$.
	
	These combined checks ensure that the numerical results are well converged and that truncation effects do not affect the physical conclusions presented in this paper.
	
	\section{Summary and conclusions}
	We demonstrate the formation of entangled Schr\"odinger cat-like states (SCLSs) in both fundamental and second-harmonic modes during degenerate spontaneous parametric down-conversion (SPDC) with quantum-depleted pumps, under both non-dissipative ($\gamma_{1,2}=0$) and dissipative ($\gamma_{1,2}=0.15$) regimes.
	At $\tau=0.38$, Wigner functions reveal clear SCLS formation with mode $\hat{a}_1$ exhibiting $N_1\approx25.28$ and mode $\hat{a}_2$ showing $N_2\approx7.35$ in the non-dissipative case, while maintaining non-Gaussianity (Wigner negativity) under dissipation. The dissipative regime introduces odd photon numbers into the SCLSs (see Figs.~\ref{fig:fig3}(c)-\ref{fig:fig3}(d)).
	
	These states demonstrate significant squeezing in quadrature components: $4.03$ dB ($\Delta^2 p_1 \approx 0.1976$) for mode $\hat{a}_1$ and $4.04$ dB ($\Delta^2 x_2 \approx 0.1970$) for mode $\hat{a}_2$, along with super-Poissonian statistics confirmed by Fano factor analysis. The super-Poissonian character ($FF \gg 1$) enhances correlation-based imaging through photon bunching \cite{a2017}, making these states valuable for quantum sensing \cite{Salykina2023,Singh2024} applications, particularly in Mach-Zehnder interferometry \cite{Shukla2023,Toscano2006,Salykina2023,Singh2024} and optical qubit encoding \cite{Yamamoto2020,Lloyd1999}.
	
	Odd SCLSs can be generated by changing the initial state to $|\psi_0\rangle = |1\rangle_1 \otimes |\alpha_{20}\rangle_2$, producing only odd photon numbers in mode $\hat{a}_1$: $e^{-it\hat{H}_{int}} |1\rangle_1 \otimes |\alpha_{20}\rangle_2 \approx |1\rangle_1 \otimes |\alpha_{20}\rangle_2 -itg \alpha_{20}\sqrt{6}|3\rangle_1 \otimes |\alpha_{20}\rangle_2$. The analysis for odd SCLSs follows the same methodology as for even states and is omitted here. Recent attempts to find optimal generation timing for SCLSs remain an active research focus \cite{Gorshenin2024}.
	
	Extending our analysis to the non-degenerate SPDC (NDSPDC) process, we find that a fully quantized pump similarly leads to non-Gaussian state formation. The pump mode develops SCLS features, while the signal mode exhibits a non-Gaussian structure distinct from the thermal state predicted by the semiclassical approximation. Notably, the tripartite NDSPDC system exhibits a higher-dimensional entanglement structure, with a Schmidt number of \(K^{(\mathrm{ND})} \approx 10.38\), compared to the bipartite entanglement (\(K \approx 1.93\)) of the degenerate process.
	
	A key finding is the formation of entanglement between the modes, a direct consequence of the fully quantized treatment of the pump. This entanglement, combined with the non-Gaussian characteristics of the SCLSs, creates a powerful resource for quantum technologies. Furthermore, the relationship between the observed non-classicality and Bell-type non-classicality, which generally differ in nature \cite{Volovich2016}, presents an interesting avenue for further investigation.
	
	These states are particularly promising for enhancing the sensitivity of Mach-Zehnder interferometers \cite{Shukla2023,Toscano2006,Salykina2023,Singh2024} and for applications in correlation-based imaging. From a practical perspective, these entangled non-Gaussian states can be generated in standard SPDC setups using periodically poled nonlinear crystals \cite[see][and references therein]{Belinsky2018}, which requires high pump conversion efficiencies (approximately 63.2\% for the degenerate case and similarly high for NDSPDC). Their verification via homodyne tomography \cite{Paul2024} is feasible and offers an advantage over $\chi^{(3)}$-based schemes due to lower pump intensities.
	
	Our findings not only contribute to the toolbox of continuous-variable quantum information processing \cite{Holevo2019} but also open new avenues for exploring the interplay between entanglement, non-Gaussianity, and non-classicality in fundamental quantum optical processes.
	
	\section*{Acknowledgements}
	We thank Prof. Anatoly V. Masalov for discussions on the obtained results and for providing valuable inputs. We also thank the Referee and Editor, Prof. Matteo G. A. Paris, for the insightful suggestions, including to extend our fully quantized treatment to the non-degenerate down-conversion configuration. Implementing this extension has allowed us to demonstrate the generation of non-Gaussian states and complex entanglement in a tripartite system, thereby broadening the scope of our work.
	
\bibliographystyle{plain}  

\end{document}